# Can AI Understand Human Personality? - Comparing Human Experts and AI Systems at Predicting Personality Correlations

Philipp Schoenegger[1], Spencer Greenberg[2], Alexander Grishin[2],

Joshua Lewis[3], and Lucius Caviola[4]

[1] London School of Economics and Political Science

[2] Spark Wave

[3] New York University

[4] University of Oxford

**Author Note**

Corresponding author: Philipp Schoenegger, contact.schoenegger@gmail.com

We declare the following potential conflict of interest: Two authors, Spencer Greenberg and Alexander Grishin, are the creators of PersonalityMap. Spencer Greenberg is also the founder of GuidedTrack.com and Positly.com.

We thank Xiaoyu (Victoria) Ouyang for research assistance in ensuring computational reproducibility.

**Abstract**

We test the abilities of specialised deep neural networks like PersonalityMap as well as general LLMs like GPT-4o and Claude 3 Opus in understanding human personality. Specifically, we compare their ability to predict correlations between personality items to the abilities of lay people and academic experts. We find that when compared with individual humans, all AI models make better predictions than the vast majority of lay people and academic experts. However, when selecting the median prediction for each item, we find a different pattern: Experts and PersonalityMap outperform LLMs and lay people on most measures. Our results suggest that while frontier LLMs' are better than most individual humans at predicting correlations between personality items, specialised models like PersonalityMap continue to match or exceed expert human performance even on some outcome measures where LLMs underperform. This provides evidence both in favour of the general capabilities of large language models and in favour of the continued place for specialised models trained and deployed for specific domains.

**Introduction**

Even though modern artificial intelligence (AI) is fundamentally distinct from humans and human intelligence, it is able to learn about human psychology. But how well do modern AI systems understand human personality? In this study, we test how accurately a set of AI models can predict the associations between human personality traits. We test whether AIs are better at this than lay people or even psychology experts. Additionally, we compare the ability to predict personality associations between generalised language models (like ChatGPT), a specialised AI system (PersonalityMap - from https://personalitymap.io) trained specifically with empirical personality data, and human comparison groups.

Modern large language models (LLMs), due to the attention mechanism (Jumper et al. 2021; Varadi et al. 2022) of the transformer architecture (Vaswani et al. 2017), have shown strong performance in a large number of diverse domains, such as marketing (Fraiwan & Khasawneh 2023), teaching (Sallam et al. 2023), programming (Bubeck et al. 2023), medicine (Nori et al. 2023; Han et al. 2023), and legal reasoning (Katz et al. 2023). Perhaps the most impressive aspect of their performance is that a single model can excel at such a distinct set of tasks, often at or above human-level and sometimes matching or even exceeding human experts at some tasks. However, there remain numerous academic and practical use cases where more specialised deep neural networks, almost always significantly smaller, continue to be used, like in the contexts of ultrasonography diagnosis (Qi et al. 2019), materials science (Qu et al. 2023), or athletics (Szijarto et al. 2023). These specialised models often outperform due to context specific requirements, direct training on the relevant goal, and the ability to leverage high quality, proprietary data (e.g., Berman et al. 2000).

In this paper, we provide a direct test of this potential divergence between types of modern artificial intelligence models. Specifically, we test the ability of two frontier LLMs, a specialised model, lay people, and academic experts to predict the relationship between two survey questions (called 'items') drawn from the personality psychology literature, testing how well these models can understand the details of human personality. For example, we may ask the model or human participant to predict the relationship (in correlation coefficient) between the following two items: "I seek to influence others" and "I enjoy interactions less than others". The former may be related to Machiavellianism (Monaghan et al. 2020), a manipulative personality trait of the dark triad (Paulhus & Williams 2002), while the second may be best understood as relating to introversion (Walker 2020), where those who score higher on introvert measures may agree with the statement above more frequently than those who do not. Predicting the relationship between these two items may be possible by direct recourse to the academic literature, though the specific pairs queried are at best only indirectly studied, like in this case (Allsopp et al. 1991). Some parts of these relationships may also be gleaned by a casual understanding of human psychology or other personal experiences. This makes this data an interesting test case, as generalised models may be able to grasp a large portion of these relationships directly, though specialised models may be even more performant due to having been trained on data like this, though whether this is the case remains an open question. For all these comparisons, we provide human lay and human expert comparisons to properly contextualise model performance in relative terms.

There is not much previous academic work that has compared lay people, experts, and machine predictions on the relationship between psychometric items from personality research. Some related work (Jayaratne & Jayatilleke 2020) has looked at Random Forest predictions of personality traits based on written interview data. Their model reliably inferred HEXACO personality traits with average correlations of between 0.31 and 0.39 depending on modelling choices. Recent work in the context of expert predictions of long-run RCT effects has found that while academic experts predict effect sizes of studies better than laypeople, they fail to pass simple benchmarks (Bernard & Schoenegger 2024), suggesting that humans generally struggle to predict outcomes of academic studies and results.

In more recent relevant literature, Hommel & Arslan (2024) introduce SurveyBot3000, a fine-tune of the sentence transformer all-mpnet-base-v2. In their analysis of data drawn from Bainbridge et al. (2022), they show that their approach can accurately predict correlations between survey items. Wulff & Mata (2023), using fine-tuned embedding models, ranging from the BERT-type MPNet to OpenAI's text-embedding-3-large and others, also manage to reconstruct relations between internal consistency measures of different psychological measures and enable prediction of empirical relations between these.

Our work contributes to this literature by testing frontier LLMs against a specialised model in their ability to predict correlations between human responses to personality items. We compare their predictive capabilities with those of a set of lay people and academic experts in psychology, which are demanding benchmarks. This enables us to test the machine capabilities of advanced, modern, and widely available models like GPT-4o with that of people holding or working towards a PhD in psychology or related disciplines. As such, we are thus able to show not just in absolute terms how well different machine approaches can predict the relationships between personality items and understand human psychology, but also how this compares to the benchmark abilities of lay people and experts.

**Hypotheses**

In this paper, we test two types of question: First, how do single machine approaches (LLMs, specialised deep neural networks) rank in the distribution of individual humans (lay people and academic experts) in the accuracy of their predicted correlations between psychometric personality items? Second, how do all the machine and human approaches compare when we use the median estimates from each approach (for a given item correlation)? For each type of question, we conduct several preregistered analyses. This allows us to better understand the distinct capabilities and potential applications of machine and human approaches.

Our first research question is about how machine approaches perform in the distribution of human predictions. We approach it in two ways. First, we test whether the machine approaches have a better or worse average error (over all items) than the median lay person and the median expert (ranked based on the average error for the subset of questions they answered). Formally, we test the preregistered null hypothesis:

*Null Hypothesis 1a: The machine approaches do not outperform or underperform the median percentile rank of lay humans and experts.*

Second, we also directly compare each individual's predictions one-to-one with a machine's to investigate whether machine models beat most humans on the questions that they answered. That is, for each human, we count each machine approach as having "won" if it has a lower average error than the human in more than half of the items that the humans made estimates for.[1] Formally, we test the preregistered null hypothesis:

*Null Hypothesis 1b: The machine approaches do not win more or less than 50% of individual comparisons to lay humans or experts.[2]*

These two hypotheses allow us to test the machine approaches on an individual level, i.e., how they would perform in a human context where a single model is introduced and pitted against a set of individual humans (both lay people and academic experts). This approach might provide results most relevant to applied contexts, as individual models may easily be integrated into a group of individual humans.

---

[1] Note, because this comparison is with the median error, human populations do not benefit from error cancellation from aggregation, as they would if we took the error of the median estimate as we do in the second research question.

[2] The wording of NH1b differs from our preregistered wording. This is because our preregistered wording was almost indistinguishable from NH1a. The analysis remains wholly unchanged, in that there are no deviations from our protocol. We have made this change solely to improve comprehensibility of our hypothesis.

Our second set of questions compares the accuracy of all the approaches by taking the median prediction for each item within each condition before comparison. We then analyse the differences in prediction error, prediction correlation, and bucketised prediction error between the conditions. These individual scores indicate how well the different approaches work as an aggregate. We test the following three preregistered null hypotheses.

*Null Hypothesis 2a: There is no difference in prediction error between the conditions.*

*Null Hypothesis 2b: There is no difference in prediction correlation between the conditions.*

*Null Hypothesis 2c: There is no difference in bucketised prediction error between the Conditions.*

This set of hypotheses lets us test the predictive capability of all approaches as an aggregate. Specifically, this allows us to benefit from the error-cancellation that aggregating individual predictions brings with it, called the 'wisdom of the crowd' effect: This 'wisdom of the crowd' effect has been documented in both human (Budescu & Chen 2015) and machine (Schoenegger et al. 2024) contexts. In our case, we aggregate across 3 items for the LLM conditions and over at least 16 predictions in each of the human conditions on each item. We then test these aggregate values on a set of three distinct outcome variables. Overall, this approach provides general results with respect to the capabilities of a given model or human population.

## Methodology

Our study collects data from five different sources. All different sources provide correlations between sets of item pairs drawn from the personality psychology literature First, we collect predictions from a layperson population recruited via Positly, an online research subject aggregation platform. Second, we use academic experts (graduate students or PhD holders in psychology or related disciplines) as our second data source. For our third and fourth sources, we query two frontier LLMs (GPT-4o and Claude 3 Opus). Fifth, we use a proprietary deep neural network called PersonalityMap as our last data source. We preregistered our data collection and analysis plans on the Open Science Framework.[3]

**Data**

For our data set, we draw on 249 pairs of personality psychology items taken from the SAPA Personality Inventory (Condon, n.d.). For example, one such item pair might be "I am an extraordinary person" and "I am easily discouraged". Overall, our data set has 103 unique items. We sampled one third of pairs to have an empirical correlation of less than -0.2, a third to have a correlation between -0.2 and 0.2, and a third to have a correlation above 0.2, in order to ensure that our test captured various types of correlations as opposed to focusing on one, e.g., small-to-nonexistent correlations between random items. For a full list of item pairs, see Appendix A.

**Sample Size Justification**

To arrive at our sample size and data numbers, we conducted the following preregistered sample size justification. Using a standard small-to-medium effect size of f=0.225 as our smallest effect size of interest for the one-way ANOVA used in our aggregate-level analysis, this would require a total of 245 participants at 80% power and an alpha level of 0.05. As our aggregate-level analysis is conducted at the question level, this corresponds to a total of 245 question pairs. We use 249 question pairs to meet this target. In order to ensure at least three predictions per item for meaningful aggregation and to account for participants answering less questions than expected, we conducted a Monte Carlo simulation, where the results indicate we need to recruit at least 119 participants. To account for potential drop-out for a variety of reasons (wrong expert status, incomplete surveys, etc, we were aiming to recruit 250 participants in the lay participant condition to account for partial completions and to ensure that our study is well-powered. For the expert condition, we aimed to recruit 200 participants who are experts in psychological research/behavioural science (graduate students or PhD holders in these fields). Our samples were willing to answer more questions than anticipated, with the median number of responses for both conditions being 30, and at the lowest number of responses for an item pair being 16 for the lay condition and 18 for the expert condition. For the LLM conditions, we collected three runs for each item. For PersonalityMap, we collected only a single prediction as the model is deterministic.

---

[3] https://osf.io/g4qm9/?view_only=633604f20ba3451cbac9852a8e9e68c0

**Participants**

We recruited a total of 254 participants via the Positly.com platform. The mean age was 46.35 years (SD=11.83), with 56% of participants identifying as male. Participants were paid a total of $1.80 each for participation (averaging to approximately $8.40 per hour). The survey included a total of 30 randomly selected item pairs that participants were asked to evaluate the relationship between. They entered their prediction on a slider ranging from -1 to +1 in increments of 0.02. Prior to their correlation estimation task, all participants completed a short introduction to correlations that included some theoretical explanation as well as examples of large, small, and zero correlations. Participants had to correctly answer some questions about correlations (to demonstrate they understood the concepts) before proceeding, though they could try as many times as they liked. The study was implemented using the GuidedTrack.com study creation platform. The study can be viewed precisely as participants saw it here https://www.guidedtrack.com/programs/18blaeo/preview and the study code can be found and copied here: https://www.guidedtrack.com/programs/27606/edit (available after creating a free login). See Figure 1 below for a graphical illustration of the task presented to lay participants and academic experts.

**Figure 1**

*Task Example*

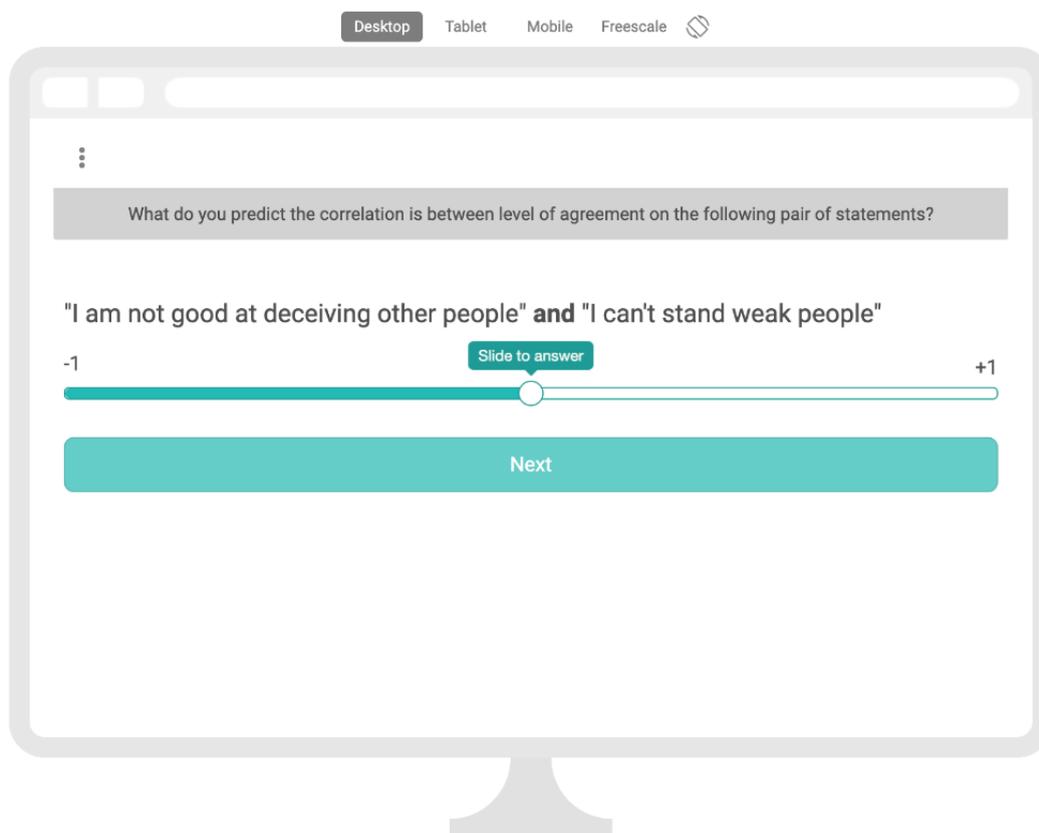

Additionally, we recruited a total of 272 participants from a number of academic email lists like SJDM as well as social media such as LinkedIn and X/Twitter. The mean age of our sample was 33.86 years (SD=8.12), with 52% of participants identifying as men. Participants were academics in psychology/behavioural science, which we defined as graduate students, postdocs, or professors. Overall, 36% of participants were professors, with 20% having completed a PhD without being a professor and 44% being graduate students. All participants were presented with a total of 30 randomly selected item pairs. They entered their estimated correlations on a slider ranging from -1 to +1 in steps of 0.02. Participants were provided with the opportunity to complete the same correlation training that all lay participants had to complete, but they could easily opt out of it. Expert participants received a $5 gift card and had the option of donating the amount to charity instead. The study can be viewed precisely as expert participants saw it.[4]

**PersonalityMap**

We collected our correlation estimates for the specialised model from a proprietary model called PersonalityMap that was developed by the startup foundry Spark Wave. It is a supervised training model that uses a pair of psychometric personality items as input. A fully connected deep neural network tries to generate the correlation between them, and after a correlation has been produced, compares it with the target correlation, adjusting the weights according to a backpropagation-based algorithm. The network was trained for 501 epochs using 992,003 item pairs as the training data, with each assigned a numerical embedding vector. In addition, 51,336 item pairs were withheld from training and used as the test set for measuring performance. The momentum was set to 0.9, and the learning rate was linearly increased from zero to 4e-5 for one epoch, after which it was decreased using cosine for 500 epochs, at which point it became 4e-6. Weights of trainable connections were initialised using standard He initialisation. The hidden layer size was set to 5 layers in total. We collect the model's predicted correlations on all 249 item pairs, all of which were not part of the data set used to train PersonalityMap (i.e., no data from any of the items that were part of the 249 item pairs was used to train the model).

---

[4] https://www.guidedtrack.com/programs/18blaeo/preview?expert=1

**Large Language Models**

We collected correlation estimates from two of the most advanced large language models at the time of conducting this study: Claude 3 Opus (claude-3-opus-20240229) with a 200,000 token context window and training data up to August 2023 and GPT-4o (gpt-4o-2024-05-13) with a 128,000 token context window and training data up to October 2023. Each model was queried with an advanced prompt via the respective API. We queried each model at temperature (T) = 0. However, to account for the sparse mixture-of-experts (MoE) architecture (Artetxe et al. 2021; Fedus 2022) that is likely used in both models due to their non-deterministic output even at T=0 (Chann 2023), we queried both models three times. This allows us to partially reduce variance inherent in the batched inference approach of this architecture (Puigcerver et al. 2023) irrespective of the otherwise expected determinacy of T = 0 that would be preferable for capability assessment.

We used a zero-shot approach without in-context learning, drawing on the following prompt for runs.[5] The prompt design drew heavily on current best practices of LLM prompting, including standard chain-of-thought and step-by-step reasoning (Wei et al. 2022) to increase reasoning capabilities, while also making use of the expert-persona framing technique (Xu et al. 2023) in order to increase model confidence and reasoning complexity. This included identification of potentially relevant psychological constructs and previous literature on these. The model is also instructed to reason from least-to-most complex justifications (Zhou et al. 2022) to consider different levels of abstraction and to take a deep breath (Yang et al. 2023) to further improve general performance. Then, the prompt instructs the model to make use of tree-of-thoughts (Yao et al. 2023) and to rely on a self-consistency constraint (Wang et al.

---

[5] The prompt for Claude 3 Opus included an additional phrase for the seventh step to ensure consistent outputs: "(Predicted Correlation: XX.XX)"

2022) in its reasoning to consider alternative explanations, ensuring that considerations of no correlation at all are continually considered. Following this, we ask the model to think of reasons against its estimate (Wang et al. 2023) to further induce critical reflection before proceeding to highlight the emotional stakes (Li et al. 2023) of this task in an attempt to increase model effortfulness, which is also what we aim to elicit with a tipping reminder and further focus on personal real-world stakes. Using GPT-4's tokenizer, this prompt without specific items amounts to 671 tokens. For the full prompt, see Appendix B.

## Results

As with our research questions, we split the results section into two parts. In the first part, we treat the machine approaches as individual data points that are compared to individual lay people and academic experts. This allows us to test how well an instance of a given machine approach may work compared with actual academic experts in the field or lay people. In the second part, we use simple aggregation for all conditions, using the median prediction at each question pair. This evens the playing field and allows the human approaches to draw on the wisdom of crowds of a set of heterogeneous respondents. Unless specifically indicated, all analyses below were preregistered on the Open Science Framework.[6] For an example output of the two LLMs, see Appendix C.

**Individual Comparisons**

To address our first research question (hypotheses 1a and 1b) of testing how well individual models do against individual humans, we randomly select one of the three LLM instances for both GPT-4o and Claude 3 Opus for both analyses. Then, we compute the absolute prediction error between the predicted correlation and the empirical correlation between the two items with lower scores indicating higher accuracy.

---
[6] https://osf.io/g4qm9/?view_only=633604f20ba3451cbac9852a8e9e68c0

In Table 1, we provide descriptive results of the prediction errors, reporting means and standard deviations of average scores at the predictor level (as opposed to means and standard deviations of individual, predictor-question-level errors that contribute to those averages). The machine average prediction errors are calculated across all items (separately for each model run), whereas each human's average prediction error is calculated across only the subset of items for which the human made estimates. We then average the errors for each person or model run (i.e, separate averages for each of the first, second, and third model run for the LLMs or one average for the single model run for personality map). In Table 1, we report the mean and standard deviation across these averages for each condition.[7]

**Table 1**

*Summary Statistics of Averaged Prediction Errors*

| Condition | Mean | Standard Deviation | Minimum | Maximum |
| --- | --- | --- | --- | --- |
| Lay | 0.29 | 0.09 | 0.11 | 0.65 |
| Expert | 0.20 | 0.09 | 0.08 | 0.86 |
| GPT-4o | 0.14 | <0.01 | 0.14 | 0.15 |
| Claude 3 Opus | 0.11 | <0.01 | 0.11 | 0.11 |
| PersonalityMap | 0.07 | - | - | - |

*Note.* Mean of average scores of individual humans or model runs. For lay people and experts, we calculate the average prediction errors across questions for each individual, based on the subset of questions which the relevant individual predicts. Then, treating the average prediction error for each individual as a separate observation, we calculate the mean and standard deviation across those average prediction errors. For GPT-4o and Claude 3 Opus, we calculate the average prediction error across questions separately for each of the

---

[7] This difference does not confer machines with a wisdom-of-crowds based advantage because we are computing the average of the errors, not the error in the average. However, it is also possible that, by a fluke of what items the humans were randomly assigned to predict, some item correlations might arise more in the human sample than in the machine sample (although this source of noise is accounted for in the confidence intervals that we report below). Finally, it is possible that some small minority of item correlations might be outliers in how easy they are to predict, bringing down the machine average, but only affecting a few humans for whom those questions happened to be in their prediction set. If so, then machines might have a lower observed average error than the majority of humans without having better-than-median true prediction accuracy. Our win-rate analyses deal with this issue by using the exact same prediction items for both humans and machines.

first, second, and third runs of the relevant model, where each run includes one prediction error for every question. Then, treating the average prediction error for each of the three runs as a separate observation, we calculate the means and standard deviations across those three average prediction errors. Both LLM conditions have minimal variability due to temperature=0 and low run count (n=3). PersonalityMap is deterministic and only has a single run. Lower scores indicate higher accuracy.

We start our main analysis by testing *Null Hypothesis 1a: The machine approaches do not outperform or underperform the median percentile rank of lay humans and experts.*

For this hypothesis, we preregistered to determine the rank of machine approaches in the following way. First, we take each individual's (or model's) predicted correlations across all item pairs and calculate the average prediction error for that individual (or model). Note, to calculate the LLM averages, we randomly select one of the LLM's three estimates for this analysis to avoid conferring LLMs with a wisdom-of-crowds avantage.

For each of the three machine approaches, we calculated the percentile rank of their average estimate with respect to both the lay and the expert human populations. Then, to test our Null Hypothesis 1a, we compute 95% confidence intervals via bootstrapping, resampling with replacement at the question level over 10,000 iterations. We find that all three machine approaches significantly outperform the preregistered 50% baseline, ranging from a rank of 70.22 (95% CI [60.66, 81.62]) for GPT-4o in comparison with the academic experts to 100 (95% CI [100, 100]) for PersonalityMap with respect to the lay population, see Table 2 for full results and Figure 2 for a visualisation of the same analysis. This allows us to reject our first null hypothesis (1a).

**Table 2**

*Percentile Rank*

| Model | Comparison | Percentile Rank CI |
| --- | --- | --- |
| GPT-4o | Lay Individuals | 95.67 [93.31, 98.03] |
|  | Expert Individuals | 70.22 [60.66, 81.62] |
| Claude 3 Opus | Lay Individuals | 100 [98.82, 100.00] |
|  | Expert Individuals | 95.22 [87.50, 98.16] |
| PersonalityMap | Lay Individuals | 100 [100.00, 100.00] |
|  | Expert Individuals | 100 [100.00, 100.00] |

Note. Each machine approach was compared individually to each of the human populations.

**Figure 2**

*Density Histogram of Participant Prediction Errors*

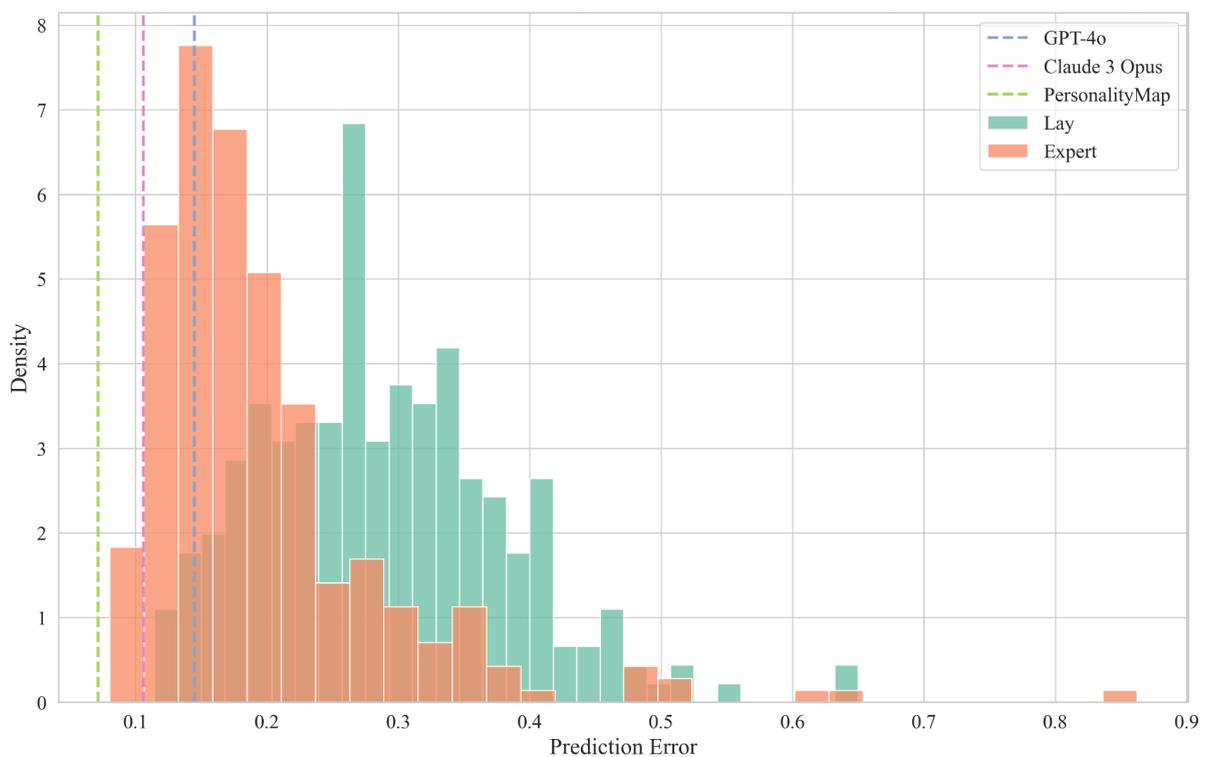

Note. Density histogram, with vertical lines showing the average error across all items for GPT-4o and Claude 3 Opus (be§cause we elicited three estimates from each model for each item, we randomly select one model for all items to calculate the average) as well as for the single PersonalityMap instance, with each model showing lower error than the majority of both human populations. Human prediction errors are calculated for each individual person separately (but averaged across all that person's predictions).

In Figure 3, we also plot the prediction errors of each prediction made within each condition without aggregating at the person-level, showing the differences in distribution of all predictions. One take-away from this plot is that PersonalityMap has a heavily concentrated number of prediction errors close to 0, while all other conditions are more widely dispersed, with human distributions having more predictions with higher errors.

**Figure 3**

*Histogram of Prediction Errors by Condition*

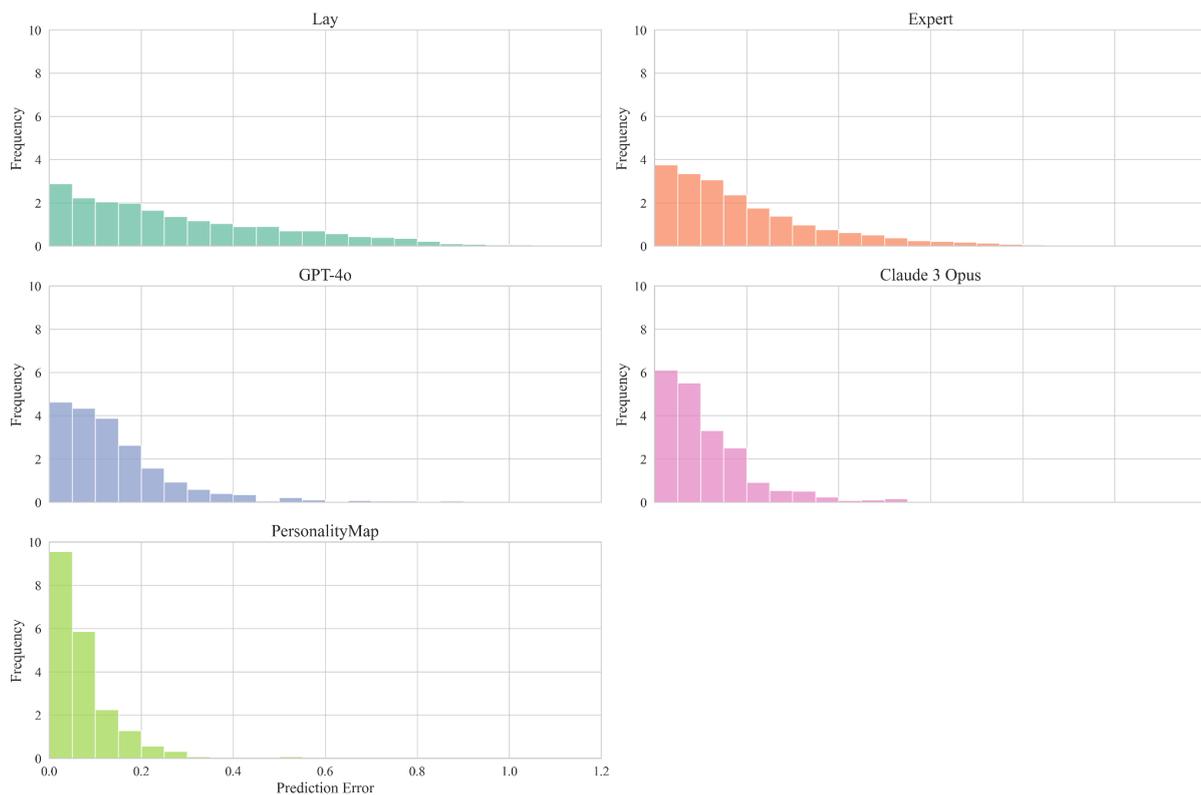

*Note.* Histogram with prediction errors that are calculated at item level, showing that machine predictions are more likely to be accurate than human ones. To improve visual clarity we do not plot errors greater than 1.2, which accounts for 0.13% of errors for the lay sample and 0.04% for the expert sample.

Next, we test *Null Hypothesis 1b: The machine approaches do not win more or less than 50% of individual comparisons to lay humans or experts.*

To do so, we used the same randomly selected instances of GPT-4o, Claude 3 Opus, and PersonalityMap as in our previous analysis, i.e., we randomly selected one of the three LLM runs for all analyses while focusing on the single PersonalityMap run. We performed an item-by-item comparison for each human participant, considering only the questions answered by both the participant and the models. For each comparison, we identified common questions and recorded the prediction errors of the models and the human participants. A "win" for a model was defined as having a lower prediction error than the human participant on more than half of the common questions, with ties resolved in favour of the human participant. In other words, there was one observation for each participant, indicating whether the given machine approach outperformed the human for more than half of the estimates that the human made. We defined the win rate for each machine approach as the percentage of wins for each model, calculated separately for lay and expert populations. The results showed that all three models demonstrated superior performance against the lay population, each achieving a win rate of over 90%. Against the expert population, GPT-4o had a win rate of 69.85%, and PersonalityMap showed a win rate of 99.26%. All approaches are significantly different from the preregistered 50% baseline. This allows us to also reject our second null hypothesis (1b).

**Table 3**

*Win Rate*

| Model | Comparison | Win Rate | Binomial Test (p-value) |
| --- | --- | --- | --- |
| GPT-4o | Lay Individuals | 90.94% | <.001 |
|  | Expert Individuals | 69.85% | <.001 |
| Claude 3 Opus | Lay Individuals | 97.64% | <.001 |
|  | Expert Individuals | 86.40% | <.001 |
| PersonalityMap | Lay Individuals | 100.00% | <.001 |
|  | Expert Individuals | 99.26% | <.001 |

*Note.* Binomial test is conducted against the 50% baseline, showing that all machine approaches outperform this baseline and win more than half the individual matchups.

The results from the first set of analyses show that all machine approaches significantly outperform the majority of lay humans as well as academic experts. This suggests that on an individual-level comparison, LLMs as well as PersonalityMap show superior personality correlation prediction capabilities than most academic experts.

**Aggregate Comparisons**

Our second research question is to compare the performance of machine and human approaches based on a representative estimate for each item pair. Thus, the unit of observation is each of the 249 item pairs. We calculated variables measuring the prediction accuracy of a representative prediction from each approach for each of these item pairs. For the LLMs, for the representative prediction, we take the median of all three predictions on each item, while for PersonalityMap, we take the single prediction it provides per item. For both human conditions, we also compute the median on each item (after having removed all missing values, as each individual human participant only responded to a fraction of the total item pairs). This procedure leaves us with a single, representative prediction in each condition for each item. Then, we conduct further analysis from this point onward.

For *Null Hypothesis 2a: There is no difference in prediction error between the conditions*, we first tested the assumptions of normality (Shapiro-Wilk) as well as homogeneity of variances (Levene's), showing assumption violations for both, $p<0.001$. Based on these results, we conducted a Kruskal-Wallis H-test, finding significant differences between the conditions. $H(4)=90.84$, $p<0.001$. This allows us to reject Null Hypothesis 2a. See Figure 4 for a visualisation of the mean prediction error for each condition.

**Figure 4**

*Mean Prediction Error by Condition*

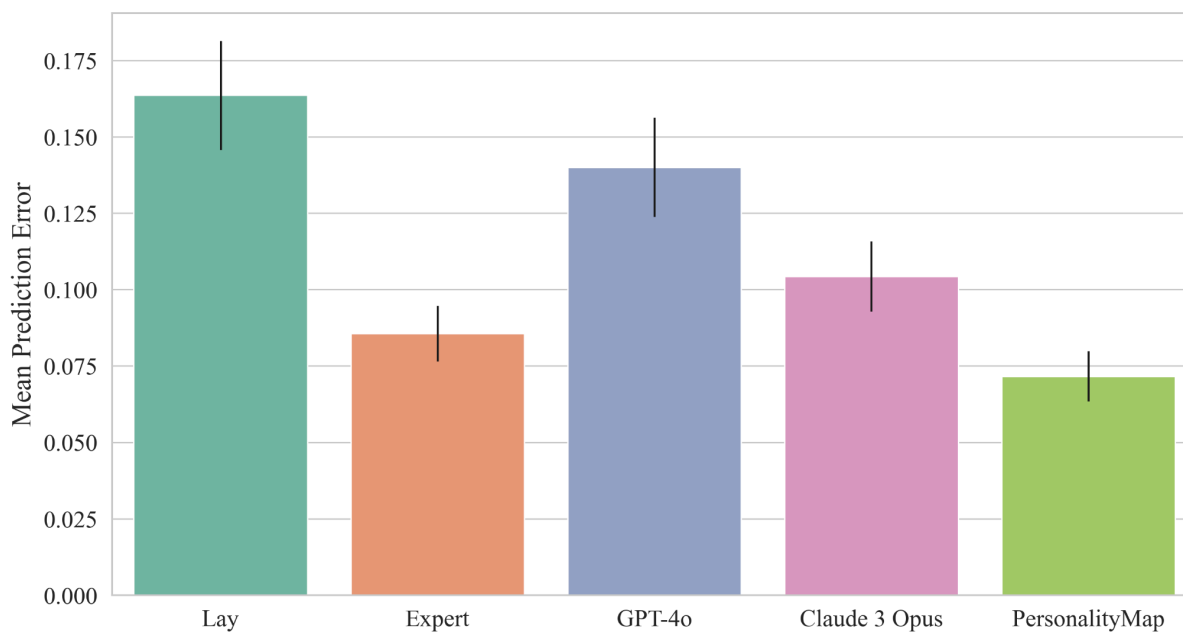

*Note.* Bar chart with 95% confidence intervals, showing that PersonalityMap and experts outperform lay people and GPT-4o. Lower bars indicate superior performance.

We then ran a Dunn's post-hoc test, employing a Bonferroni correction for multiple comparisons. We find that Claude 3 Opus outperforms both GPT-4o, p=.011, and the lay population, p<.001, but is outperformed by PersonalityMap, p<.001, with no significant differences to the academic expert sample, p=.329. PersonalityMap is similarly not significantly different from the expert population, p=.444, but outperforms all other conditions at medium-to-large effect sizes with Hedge's g between 0.41 and 0.82. GPT-4o is not statistically different from the lay population, p=1.000, significantly underperforming all other conditions at medium effect sizes with Hedge's between 0.31 and 0.66. See Table 4 for the full pairwise comparison results.

**Table 4**

*Post-Hoc Pairwise Comparisons*

| Comparison | p-value (adj.) | Hedge's g |
|---|---|---|
| Claude 3 Opus vs Expert | .329 | 0.22 |
| Claude 3 Opus vs GPT-4o | .011 | -0.31 |
| Claude 3 Opus vs Lay | <.001 | -0.49 |
| Claude 3 Opus vs PersonalityMap | <.001 | 0.41 |
| Expert vs GPT-4o | <.001 | -0.51 |
| Expert vs Lay | <.001 | -0.68 |
| Expert vs PersonalityMap | .444 | 0.20 |
| GPT-4o vs Lay | 1.000 | -0.17 |
| GPT-4o vs PersonalityMap | <.001 | 0.66 |
| Lay vs PersonalityMap | <.001 | 0.82 |

*Note.* The p-values are adjusted via the Bonferroni correction.

For *Null Hypothesis 2b: There is no difference in prediction correlation between the conditions*, we test whether the conditions differ in their correlation between the aggregate predictions and the empirical correlation values. To better allow for comparisons between correlations that are all similarly high, we normalise the correlations via the Fisher's Z transformation prior to computing 95% confidence intervals via bootstrapping with 10,000 iterations, where we resample the questions. We can reject Null Hypothesis 2b as we find differences between some conditions. For correlation coefficients and Fisher's Z values with bootstrapped 95% confidence intervals, see Table 5.

**Table 5**

*Pearson Correlations and Fisher's Z Values*

| **Condition** | **Pearson Correlation** | **Fisher's Z (95% CI)** |
|---|---|---|
| Lay | 0.88 | 1.37 (1.25, 1.50) |
| Expert | 0.90 | 1.45 (1.33, 1.57) |
| GPT-4o | 0.78 | 1.05 (0.87, 1.25) |
| Claude 3 Opus | 0.80 | 1.11 (0.96, 1.26) |
| PersonalityMap | 0.91 | 1.52 (1.35, 1.69) |

*Note.* Bootstrapping resamples the questions, showing that PersonalityMap and Expert correlations are higher than the other approaches'.

We can get a sense of which approaches statistically outperform other approaches based on whether there is overlap in their Fisher's Z 95% confidence intervals in Table 5. These confidence intervals indicate that, in terms of the relationship between predicted correlations and actual empirical correlations, PersonalityMap outperforms both GPT-4o and Claude 3 Opus, with both human conditions being indistinguishable from it. There is also evidence of human samples showing higher correlations than both GPT-4o and Claude 3 Opus, although there is some very slight overlap between the lay population's confidence interval and those of each LLM.

It is notable that this analysis is the first in which lay predictors appear to outperform LLMs. This outperformance arises in part because of the aforementioned aggregation effects, as non-aggregated correlations for the human samples would be much lower, at 0.68 and 0.70 for the lay and expert populations respectively. However, this aggregation is unlikely to be the whole explanation, as even in our analyses of aggregated prediction errors, LLMs (particularly Claude 3 Opus) appeared to outperform humans. Why might LLMs have lower average prediction errors yet also have estimates less correlated with the truth? We speculate that machine approaches are well calibrated to the magnitude of personality-item correlations in general, and thus make estimates of a similar range of magnitudes to that of the correlations we observe. This advantage reduces average errors compared to humans (e.g. a human who predicted a correlation of .99 for a true correlation of .1 and -.99 for a correlation of -.1 one would have very bad average errors compared to a machine that estimated, say, 0 and -.15 respectively). In contrast, humans might be better at anticipating the relative magnitude and direction of different types of relationships. This human advantage improves the correlation between their predictions and the truth (e.g. a human who predicted a correlation of .99 for a true correlation of .1, 0 for a true correlation of 0, and -.99 for a correlation of -.1 would have a superior correlation with the truth than a machine that gave the more-accurate-but-incorrectly-ordered predictions of 0, .15, and -.15 respectively). Interestingly, while experts vastly outperform lay people in terms of average prediction error, there is no detectable difference between experts and lay predictors in terms of the correlation of their predictions with the truth. This distinction hints that experts' advantage over lay people is more driven by appreciation of the typical magnitude of personality-item correlations rather than insight into the direction or strength of relationships. See Figure 5 for a graphical overview of these results.

**Figure 5**

*Prediction Correlation by Condition*

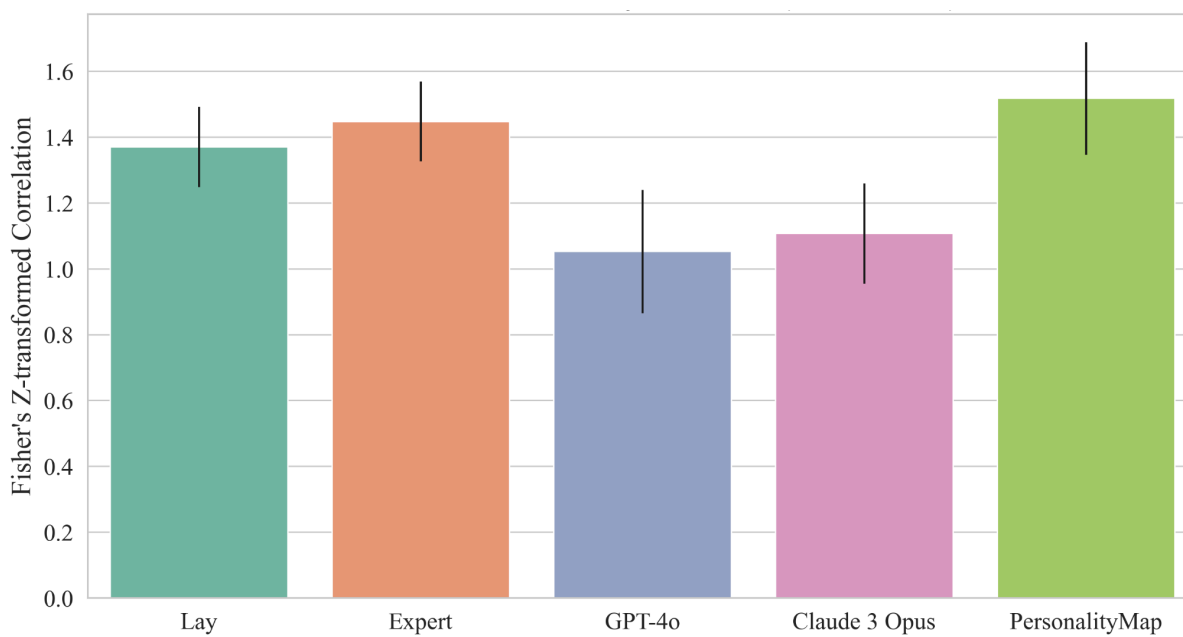

*Note.* Plot shows Fisher's Z-transformed correlations and bootstrapped 95% confidence intervals, showing that LLM predictions are less correlated with true correlations than the other conditions.

For our analysis of *Null Hypothesis 2c: There is no difference in bucketised prediction error between the conditions*, we analyse the frequency with which the different approaches' predictions fall into the correct bucket as opposed to using prediction error with the detailed empirical correlation. For this, we classified predictions as falling or not falling into the following buckets: [<-0.1; -0.1 to 0.1; >0.1].

**Table 6**

*Contingency Table*

| Condition | Correct | Incorrect |
|---|---|---|
| Lay | 196 | 53 |
| Expert | 193 | 56 |
| GPT-4o | 183 | 66 |
| Claude 3 Opus | 184 | 65 |
| PersonalityMap | 203 | 46 |

*Note.* Count of correctly and incorrectly identified correlations.

We then conduct a Chi-square test of independence, where we fail to find statistically significant differences between the conditions, $\chi^2$ (4, N = 1245) = 6.42, p = 0.170, see Figure 6. As such, we are unable to reject our Null Hypothesis 2c. See Figure 6.

**Figure 6**

*Frequency of Correct Buckets by Condition*

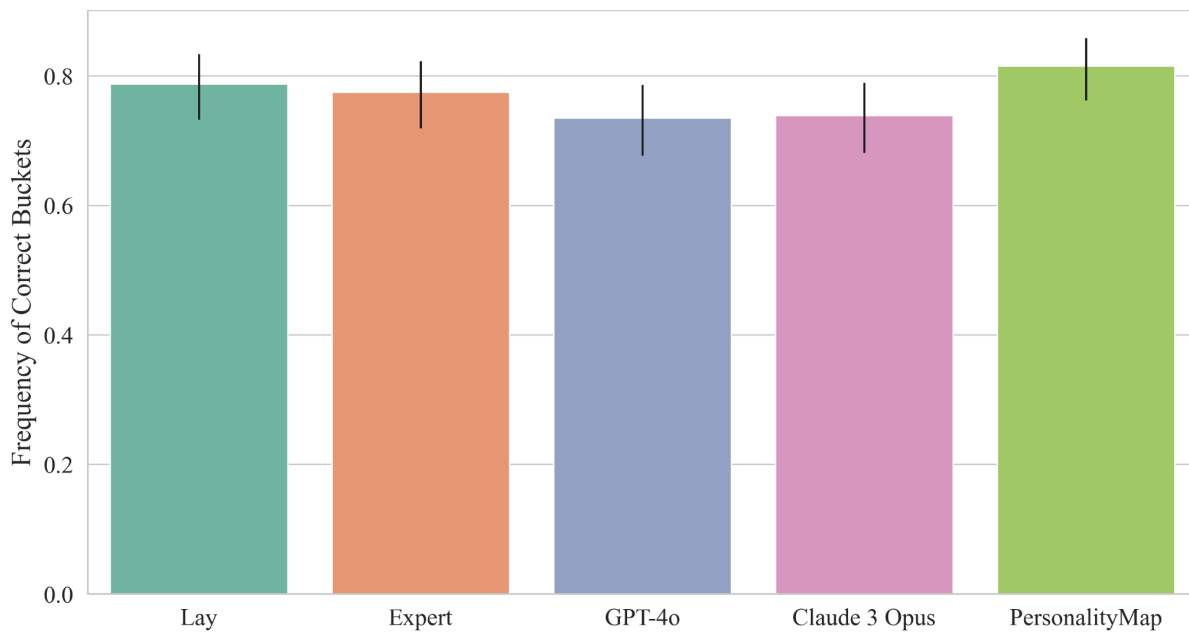

*Note.* Bar plot with 95% confidence intervals, showing no differences in correct frequency across conditions.

## Discussion

Our results suggest that current AI models are roughly as good as, if not better, than human experts in predicting correlations amongst human personality traits. Specifically, we found that AI models perform much better than the vast majority of individual lay people as well as academic experts. However, expert groups can match specialised AI model performance when their predictions are aggregated by taking the median prediction for each question, and are able to exceed LLM performance. This effect showcases the wisdom of crowds of human populations, i.e. the phenomenon where the collective judgement of a group is often more accurate than that of individual members. In other words, current frontier LLM performance (with respect to predicting personality associations) is currently somewhere between what an individual person can do and what a group of experts can do together. Crucially, we also discovered that a specialised AI system trained on personality data (i.e., PersonalityMap) performed better than generalised AI systems (i.e. current frontier LLMs), outperforming even aggregate expert estimates.

By comparing LLMs and PersonalityMap to the full set of individual lay people and academic experts, we find that all machine approaches are significantly better than the median human counterpart and win significantly more than 50% of 1-on-1 comparisons. In fact, PersonalityMap and Claude 3 Opus outperform at least 85% on both of these metrics. This suggests that when individual models are employed in a task like this, one can expect them to outperform most individual human experts that one may have consulted otherwise. This opens up a lot of potential applications of machine approaches, both for general LLMs but especially for targeted ones like PersonalityMap, with the former also being able to facilitate a back-and-forth about the area, see Appendix B, and the latter with the provision of accurate predictions that may help expert applications.

Contrary to this rather clean result, the data collected to answer our second set of research questions is considerably more mixed. For these analyses, we use one representative prediction within each condition, i.e., each condition has the median prediction per item pair. Testing for the differences between the average accuracy of these aggregated forecasts, we find that GPT-4o is not statistically different from the set of lay individuals, while Claude 3 Opus and PersonalityMap are better than both but remain statistically indistinguishable from the academic expert sample. This pattern is somewhat replicated with respect to the outcome measure of correlations between predicted and empirical correlation values, where PersonalityMap is indistinguishable from the human approaches, all outperforming the two LLMs. On our last measure of bucketised accuracy, we fail to find any differences, suggesting that what makes the best prediction methods stand out is not whether they get the direction of predictions correct, but how close they are to the actual correlation values.

Why do the LLMs appear to have a reduced advantage in the second set of research questions generally? We speculate that the main reason for these different patterns of results is that in the second set of analyses, the human estimates are boosted by the fact that they can rely on the wisdom of the crowd (Surowiecki 2012), a well-established phenomenon that shows that aggregate values can cancel out individual errors and improve predictions across a wide set of contexts. Specifically, for humans, we take the median of the humans' estimates, which would otherwise be subject to much more noise. The machine estimates are subject to less noise (zero in the case of Personality Map) and so taking the median estimate does less to improve accuracy as the aggregation occurs only over three judgments for the LLMs that were already produced with a temperature setting at 0, compared to at least 16 judgments for the human conditions, with median aggregation being over at least 30 judgments. The main take-away from this is that, in contrast to individual comparisons, as an aggregate, academic experts are not outperformed by the machine approaches. However, acquiring the necessary results from a human academic expert sample is expensive and time-consuming, while many machine approaches are much cheaper and easier to implement.

Although Personality Mapped performed very well on all measures, it is also notable that humans (and particularly lay people) performed relatively better vs. LLMs in terms of correlation and bucketised predictions than in terms of average prediction error. We speculate that this difference arises from the different strengths and weaknesses of the various approaches. Lay people appear to be particularly poorly calibrated as to the magnitude of correlations generally, but (when their judgments are aggregated) better than LLMs and close to experts in terms of their judgments of the relative strength and direction of specific correlations. In contrast, LLMs appear weaker than (aggregated) experts and lay people in terms of their judgments of the relative strength and direction of specific correlations, but superior to both human populations in their calibration to the typical magnitudes of correlations generally in our sample. These differences in strengths and weaknesses are somewhat intuitive. We might expect lay people to have very good instincts for which kind of items might draw similar responses from a human responder. Yet, we might expect large language models, and to some extent experts, to have been exposed to far more examples of similar correlations, lending them a considerable advantage in judging their magnitude. Future research could explore these conjectures more, and perhaps test how much human judgments can be improved by exposure to examples of similar correlations.

Artificial intelligence systems like PersonalityMap or GPT-4o, which can outperform even human experts on some measures of predicting psychological facts about humans, open an intriguing possibility for the future of social science research. To make an analogy, imagine if biologists could only ever conduct research in human bodies (in vivo) without the ability to do experiments in vitro. Test tube experiments allow for much faster iteration than is possible with direct human experiments (though of course, preliminary test tube results must ultimately be confirmed in humans). But no such in vitro approach to psychology experimentation has existed, until now. But now, it may be possible for researchers to generate new hypotheses and conduct preliminary tests of hypotheses before conducting a single human experiment, which may dramatically accelerate the speed of research (Manning, Zhu, & Horton 2024). Of course, as with biological experiments in test tubes, before the research is finished, the findings must be confirmed in real humans to make sure they apply. Still, if early pilot studies could be replaced with queries to machine learning algorithms, it seems possible that the research process could be substantially accelerated.

On the other hand, there are still many limitations of AI systems for personality research. While the website hosting PersonalityMap (https://personalitymap.io/) allows researchers to see predictions about the correlations between agreement to any pair of statements, as well as the ability to control for variables and conduct simulated linear regression between an independent variable and multiple dependent variables, much of what social science is interested in is causal relationships, and this technology only allows associations to be studied. Additionally, while such a system can, in theory, already make predictions about any statements whatsoever, the accuracy of such a specialised system is likely to suffer dramatically for statements that are very different from any seen during training (a fundamental challenge for machine learning models generally). Future research will also be needed to quantify the uncertainty in these predictions so that models can make researchers aware of when the predictions are more reliable and when they are less trustworthy. Finally, current incarnations of this technology only include linear relationships (represented by correlations). Perhaps future versions will also enable the modelling of non-linear relationships as well.

**Limitations**

One potential limitation of our study is that the data used, the SAPA Personality Inventory (Condon, n.d.), may be part of the training data for both LLMs, GPT-4o and Claude 3 Opus. While we can be sure that this is not the case for PersonalityMap, as the studied item pairs were not part of the training data, it is possible that they have been part of the LLM training data which may thus overstate their ability to predict correlations between the sets of personality items. However, we want to point out that we were not able to find these individual correlations publicly available.

A second limitation is that we did not incentivise either of our human condition's responses for accuracy. This may reduce their performance, which is something that future research can test.

# Appendix

## Appendix A

*Full Item List*

| Pair | Item 1 | Item 2 |
| --- | --- | --- |
| 1 | I am interested in people | When with a group, I have difficulties selecting a good topic to talk about |
| 2 | I bottle up my feelings | I don't finish the things that I start |
| 3 | I do things by the book | I don't worry about political and social problems |
| 4 | I make a fool of myself | I am always ready to start afresh |
| 5 | I like to do frightening things | I am sensitive to the needs of others |
| 6 | I don't know why I do some of the things I do | I speak softly |
| 7 | I make rash decisions | I see that rules are observed |
| 8 | I believe that others have good intentions | I feel healthy and vibrant most of the time |
| 9 | I don't like the idea of change | I do not enjoy going to art museums |
| 10 | I like to be thought of as a normal kind of person | I do things by the book |
| 11 | I love luxury | I put on a show to impress people |
| 12 | I know how to captivate people | I am exacting in my work |
| 13 | I dislike people who don't know how to behave themselves | I try to avoid complex people |
| 14 | I try to avoid complex people | When with a group, I have difficulties selecting a good topic to talk about |
| 15 | I will not probe deeply into a subject | I don't worry about things that have already happened |
| 16 | I am always ready to start afresh | I make people feel at ease |
| 17 | I would like for other people to be afraid of me | I am an extraordinary person |
| 18 | I can't stand weak people | I am a creature of habit |
| 19 | I am not good at deceiving other people | I can't stand weak people |
| 20 | I do things in a half-way manner | I make a fool of myself |
| 21 | I believe that I am better than others | I am sensitive to the needs of others |
| 22 | I have some bad habits | I carry out my plans |
| 23 | I like to get lost in thought | I need a creative outlet |
| 24 | I would like for other people to be afraid of me | I am a talkative person |
| 25 | I am easily discouraged | I stumble over my words |
| 26 | I am considered to be kind of eccentric | I put on a show to impress people |
| 27 | I am afraid to draw attention to myself | I laugh a lot |
| 28 | I seek to influence others | I love a good fight |
| 29 | I do too little work | I feel that others misunderstand me |
| 30 | I am not good at deceiving other people | I am considered to be kind of eccentric |
| 31 | I make rash decisions | I seldom get mad |
| 32 | I warm up quickly to others | I stumble over my words |
| 33 | I don't finish the things that I start | I carry out my plans |
| 34 | I often feel uncomfortable around others | I feel healthy and vibrant most of the time |
| 35 | I don't know why I do some of the things I do | I like to take it easy |
| 36 | I do things by the book | I do things according to a plan |
| 37 | I don't know why I do some of the things I do | I can be stirred up easily |
| 38 | I enjoy feeling 'close to the earth' | I am always ready to start afresh |
| 39 | I am a talkative person | I experience my emotions intensely |
| 40 | I jump into things without thinking | I feel that others misunderstand me |
| 41 | I am easily discouraged | I don't worry about things that have already happened |
| 42 | I like to get lost in thought | I love to eat |
| 43 | I enjoy interactions less than others | I am easily discouraged |
| 44 | I am easily distracted | I stick to my chosen path |
| 45 | I don't finish the things that I start | I am easily discouraged |
| 46 | I stick to my chosen path | I am mainly interested in money |
| 47 | I am interested in people | I am a very private person |
| 48 | I enjoy interactions less than others | I work hard |
| 49 | I am interested in people | I often feel uncomfortable around others |
| 50 | I like to do frightening things | I feel healthy and vibrant most of the time |
| 51 | I keep my cool | I do things without thinking of the consequences |
| 52 | I do things in a half-way manner | I do things according to a plan |
| 53 | I amuse my friends | I am sensitive to the needs of others |
| 54 | I like to be thought of as a normal kind of person | I act without thinking |
| 55 | I believe that I am better than others | I do things by the book |
| 56 | I am skilled in handling social situations | I don't finish the things that I start |
| 57 | I am skilled in handling social situations | I carry out my plans |
| 58 | I am interested in people | I rarely get caught up in the excitement |
| 59 | I am an extraordinary person | I am a very private person |
| 60 | I keep my cool | I can be stirred up easily |
| 61 | I love luxury | I only feel comfortable with friends |
| 62 | I make rash decisions | I am exacting in my work |
| 63 | I am considered to be kind of eccentric | I stick to the rules |
| 64 | I seek to influence others | I take control of things |
| 65 | I see that rules are observed | I seldom get mad |
| 66 | I seek to influence others | I rebel against authority |
| 67 | I believe that others have good intentions | I carry out my plans |
| 68 | I like to get lost in thought | I feel others' emotions |
| 69 | I believe that others have good intentions | I enjoy interactions less than others |
| 70 | I bottle up my feelings | I stick to the rules |
| 71 | I don't finish the things that I start | I feel that others misunderstand me |
| 72 | I am easily discouraged | I laugh a lot |
| 73 | I don't like the idea of change | I enjoy interactions less than others |
| 74 | I am often worried by things that I said or did | I don't worry about things that have already happened |
| 75 | I am exacting in my work | I do too little work |
| 76 | I only feel comfortable with friends | I stumble over my words |
| 77 | I feel that others misunderstand me | I have often gone against my parents' wishes |
| 78 | I am skilled in handling social situations | I laugh a lot |
| 79 | I suspect facial expressions show when sad | I don't know why I do some of the things I do |
| 80 | I am a goal-oriented person | I make people feel at ease |
| 81 | I like to take it easy | I work hard |
| 82 | I know how to captivate people | I often feel uncomfortable around others |
| 83 | I am a talkative person | I only feel comfortable with friends |
| 84 | I reveal little about myself | I can't stand weak people |
| 85 | I am not interested in abstract ideas | I speak softly |
| 86 | I enjoy interactions less than others | I am considered to be kind of eccentric |
| 87 | I bottle up my feelings | I speak softly |
| 88 | I warm up quickly to others | I feel that others misunderstand me |
| 89 | I do things in a half-way manner | I am a goal-oriented person |
| 90 | I would like for other people to be afraid of me | I like to do frightening things |
| 91 | I often feel uncomfortable around others | I don't worry about things that have already happened |
| 92 | I feel that others misunderstand me | I am afraid to draw attention to myself |
| 93 | I would like for other people to be afraid of me | When with a group, I have difficulties selecting a good topic to talk about |
| 94 | I am more easy-going about right and wrong than most people | I am a very private person |
| 95 | I am a goal-oriented person | I do too little work |
| 96 | I make rash decisions | I am easily discouraged |
| 97 | I am more easy-going about right and wrong than most people | I stick to the rules |
| 98 | I dislike people who don't know how to behave themselves | I am mainly interested in money |
| 99 | I don't know why I do some of the things I do | I don't worry about things that have already happened |
| 100 | I keep my cool | I only feel comfortable with friends |
| 101 | I know how to captivate people | I am easily discouraged |
| 102 | I am skilled in handling social situations | I don't know why I do some of the things I do |
| 103 | I feel that I have a lot of inner strength | I often feel uncomfortable around others |
| 104 | I make a fool of myself | I feel that others misunderstand me |
| 105 | I am a talkative person | When with a group, I have difficulties selecting a good topic to talk about |
| 106 | I take control of things | I rarely get caught up in the excitement |
| 107 | I make rash decisions | I am afraid to draw attention to myself |
| 108 | I do things in a half-way manner | I often feel uncomfortable around others |
| 109 | I am a goal-oriented person | I see that rules are observed |
| 110 | I seldom get mad | I feel healthy and vibrant most of the time |
| 111 | I listen to my conscience | I do things without thinking of the consequences |
| 112 | I feel that others misunderstand me | I take control of things |
| 113 | I believe that others have good intentions | I rarely get caught up in the excitement |
| 114 | I don't know why I do some of the things I do | I feel that I have a lot of inner strength |
| 115 | I dislike people who don't know how to behave themselves | I enjoy interactions less than others |
| 116 | I am skilled in handling social situations | I have some bad habits |
| 117 | I believe that others have good intentions | I don't like the idea of change |
| 118 | I can be stirred up easily | I don't worry about things that have already happened |
| 119 | I know how to captivate people | I only feel comfortable with friends |
| 120 | I like to get lost in thought | I tend to vote for liberal political candidates |
| 121 | I jump into things without thinking | I act without thinking |
| 122 | I suspect facial expressions show when sad | I jump into things without thinking |
| 123 | I like to do frightening things | I work hard |
| 124 | I am a talkative person | I feel that I have a lot of inner strength |
| 125 | I am a talkative person | I love a good fight |
| 126 | I don't know why I do some of the things I do | I have some bad habits |
| 127 | I can't stand weak people | I don't worry about things that have already happened |
| 128 | I do things in a half-way manner | I feel healthy and vibrant most of the time |
| 129 | I see beauty in things that others might not notice | I do things without thinking of the consequences |
| 130 | I break my promises | I do things without thinking of the consequences |
| 131 | I like to get lost in thought | I do things by the book |
| 132 | I laugh a lot | I feel healthy and vibrant most of the time |
| 133 | I would like for other people to be afraid of me | I rebel against authority |
| 134 | I make rash decisions | I like to do frightening things |
| 135 | I make rash decisions | I act without thinking |
| 136 | I am not bothered by disorder | I need a creative outlet |
| 137 | I do not enjoy going to art museums | I feel others' emotions |
| 138 | I break my promises | I keep my cool |
| 139 | I do things in a half-way manner | I take control of things |
| 140 | I am always ready to start afresh | I like to take it easy |
| 141 | I will not probe deeply into a subject | How often do you exercise? |
| 142 | I would like for other people to be afraid of me | People have said that I sometimes act rashly |
| 143 | I like to get lost in thought | I will not probe deeply into a subject |
| 144 | I break my promises | I work hard |
| 145 | I love a good fight | When with a group, I have difficulties selecting a good topic to talk about |
| 146 | I seek to influence others | I enjoy interactions less than others |
| 147 | I break my promises | I feel that I have a lot of inner strength |
| 148 | I see beauty in things that others might not notice | I demand quality |
| 149 | I see that rules are observed | I carry out my plans |
| 150 | I reveal little about myself | I am a very private person |
| 151 | I do things in a half-way manner | I am not easily disturbed by events |
| 152 | I feel that I have a lot of inner strength | I make people feel at ease |
| 153 | I am skilled in handling social situations | I work hard |
| 154 | I don't know why I do some of the things I do | I listen to my conscience |
| 155 | I do not enjoy going to art museums | I need a creative outlet |
| 156 | I boast about my virtues | I feel others' emotions |
| 157 | I often feel uncomfortable around others | I rarely get caught up in the excitement |
| 158 | I believe that others have good intentions | I am a talkative person |
| 159 | I am easily distracted | I stumble over my words |
| 160 | I act without thinking | I put on a show to impress people |
| 161 | I suspect facial expressions show when sad | I do too little work |
| 162 | I make rash decisions | I do things according to a plan |
| 163 | I don't like the idea of change | Are you in a committed relationship? |
| 164 | I like to do frightening things | I have often gone against my parents' wishes |
| 165 | I am a talkative person | I enjoy interactions less than others |
| 166 | I keep my cool | I put on a show to impress people |
| 167 | I seek danger | How often do you exercise? |
| 168 | I rebel against authority | I am afraid to draw attention to myself |
| 169 | I don't know why I do some of the things I do | I carry out my plans |
| 170 | I try to avoid complex people | I often feel uncomfortable around others |
| 171 | I feel that I have a lot of inner strength | I act without thinking |
| 172 | I amuse my friends | I act without thinking |
| 173 | I know how to captivate people | I am more easy-going about right and wrong than most people |
| 174 | I rebel against authority | I do things without thinking of the consequences |
| 175 | I know how to captivate people | I believe that others have good intentions |
| 176 | I am exacting in my work | I see that rules are observed |
| 177 | I warm up quickly to others | I don't consider myself religious |
| 178 | I listen to my conscience | I feel that I have a lot of inner strength |
| 179 | I make a fool of myself | I can be stirred up easily |
| 180 | I do things in a half-way manner | I am exacting in my work |
| 181 | I jump into things without thinking | I rarely get caught up in the excitement |
| 182 | I suspect facial expressions show when sad | I am considered to be kind of eccentric |
| 183 | I carry out my plans | I feel that I have a lot of inner strength |
| 184 | I make rash decisions | I feel healthy and vibrant most of the time |
| 185 | I only feel comfortable with friends | I speak softly |
| 186 | I am interested in people | I cry during movies |
| 187 | I am afraid to draw attention to myself | I am easily discouraged |
| 188 | I am not interested in abstract ideas | I warm up quickly to others |
| 189 | I don't like the idea of change | I am often worried by things that I said or did |
| 190 | I would like for other people to be afraid of me | I believe that I am better than others |
| 191 | I would like for other people to be afraid of me | I am not good at deceiving other people |
| 192 | How often do you smoke? | I am an extraordinary person |
| 193 | I keep my cool | I make people feel at ease |
| 194 | I make a fool of myself | I make rash decisions |
| 195 | I am a talkative person | I rarely get caught up in the excitement |
| 196 | I am considered to be kind of eccentric | I have often gone against my parents' wishes |
| 197 | I make rash decisions | I carry out my plans |
| 198 | I carry out my plans | I act without thinking |
| 199 | I make a fool of myself | I like to do frightening things |
| 200 | I reveal little about myself | I am a talkative person |
| 201 | I love luxury | I don't like the idea of change |
| 202 | I seldom get mad | People have said that I sometimes act rashly |
| 203 | I feel that I have a lot of inner strength | When with a group, I have difficulties selecting a good topic to talk about |
| 204 | I don't know why I do some of the things I do | I am easily distracted |
| 205 | I enjoy feeling 'close to the earth' | I am easily discouraged |
| 206 | I am skilled in handling social situations | I am easily discouraged |
| 207 | I don't know why I do some of the things I do | I feel that others misunderstand me |
| 208 | I can't stand weak people | I take control of things |
| 209 | I make a fool of myself | I don't know why I do some of the things I do |
| 210 | I make a fool of myself | I break my promises |
| 211 | I make a fool of myself | I like to be thought of as a normal kind of person |
| 212 | I do things in a half-way manner | I do things by the book |
| 213 | I do things in a half-way manner | I carry out my plans |
| 214 | I do too little work | I am sensitive to the needs of others |
| 215 | I reveal little about myself | I like to do frightening things |
| 216 | I seek to influence others | I boast about my virtues |
| 217 | I make rash decisions | I am sensitive to the needs of others |
| 218 | I do too little work | I like to tidy up |
| 219 | I am not interested in abstract ideas | I feel that I have a lot of inner strength |
| 220 | I would like for other people to be afraid of me | I make people feel at ease |
| 221 | I am an extraordinary person | I am easily discouraged |
| 222 | I love a good fight | I do things without thinking of the consequences |
| 223 | I can be stirred up easily | I am often worried by things that I said or did |
| 224 | I am skilled in handling social situations | I rarely get caught up in the excitement |
| 225 | I am considered to be kind of eccentric | I don't worry about things that have already happened |
| 226 | I have some bad habits | I seek danger |
| 227 | I seldom get mad | I experience my emotions intensely |
| 228 | I will not probe deeply into a subject | I feel healthy and vibrant most of the time |
| 229 | I like to get lost in thought | I like to be thought of as a normal kind of person |
| 230 | I listen to my conscience | I make people feel at ease |
| 231 | I can't stand weak people | I laugh a lot |
| 232 | I break my promises | I take control of things |
| 233 | I am not interested in abstract ideas | What is your weight in kilograms? |
| 234 | I seldom get mad | I feel that others misunderstand me |
| 235 | I seek to influence others | I put on a show to impress people |
| 236 | I like to visit new places | I do things according to a plan |
| 237 | I am passionate about causes | I am often worried by things that I said or did |
| 238 | I like to do frightening things | I do things without thinking of the consequences |
| 239 | I work hard | I stumble over my words |
| 240 | I am easily distracted | I have often gone against my parents' wishes |
| 241 | I seek danger | I have often gone against my parents' wishes |
| 242 | I am easily distracted | I am exacting in my work |
| 243 | I can't stand weak people | I tend to vote for liberal political candidates |
| 244 | I do things without thinking of the consequences | I tend to vote for liberal political candidates |
| 245 | I do things according to a plan | I experience my emotions intensely |
| 246 | I don't like the idea of change | I feel healthy and vibrant most of the time |
| 247 | I demand quality | I can be stirred up easily |
| 248 | I warm up quickly to others | I laugh a lot |
| 249 | I know how to captivate people | I love a good fight |

*Note.* Full list of item pairs used.

## Appendix B

*GPT-4o and Claude 3 Opus prompt*

You are a world-renowned expert psychologist. You have a long background in researching human psychology and are exceptional at statistical analysis, having published accurate research in the top peer-reviewed academic journals. You deeply understand how different aspects of human life and the psyche play together and are often asked to provide estimations of psychological results that you estimate correctly.

Your task is to estimate the correlation coefficient between two psychometric items. These items are sentences or parts of sentences. Their correlation can range from -1 to +1. You will provide correlation coefficients with at least two decimal points.

First, you begin by thinking step by step. Study each psychometric item in detail, examine them through the lens of the most accurate psychological theories and available data, and progressively articulate the logical steps that lead to your reasoning for potential correlations. These steps will include (but are not limited to):

    a) Identifying potentially key psychological constructs within the items.

    b) Drawing from empirical research that discusses the relationship between such constructs.

    c) Analyzing how these constructs have been statistically correlated in past research.

    d) Consider whether there is a correlation at all.

Second, you will structure your further assessment by progressing from least complex to most complex justifications. Start with fundamental psychological concepts and move up to more intricate and complex theories that require substantially more nuanced understanding.

Third, take a deep breath.

Fourth, summarise what you have learnt so far and map out different branches of reasoning paths, considering multiple potential outcomes and psychological interpretations of the correlations considered. Ensure that each branch represents a different line of thought, leading you through a decision-making process that mirrors traversing a decision tree.

Fifth, after exploring these various branches, use self-consistency to evaluate your findings. This involves reviewing the different lines of reasoning you have mapped out and identifying the most consistent outcomes across the different approaches you have investigated so far. Ensure that this cross-validation is thorough and lengthy. This should always include a detailed analysis of the size of the correlation, thinking through whether it should be large, medium, small, or non-existent, with small to non-existent correlations being the base case.

Sixth, consider three reasons against your analysis, outlining why you might be mistaken. Ensure that these reasons are fair and well-researched. Ensure that the estimates are not too extreme and generally fit within what is typically found in empirical research in this area, where many effects are small (-0.1 to 0.1).

Seventh, summarise your final estimated correlation between the two items in a step-by-step manner, building directly on your analysis above. Ensure that your final estimate is accurate and represents the most plausible estimate of a true correlation found in a human study of a United States of America (US) population (and is a single estimate, not a range, and is the only number you output). Note that this analysis is not a simple exercise but is actually extremely important to me personally and will have large impacts on my personal and professional life. We will test your answer in a real human sample, so ensure that you are correct; this matters! Recall that we are paying $20 for this!

The input will be:

Item 1: XXX

Item 2: XXX

The output will include your reasoning and the final answer:

Predicted Correlation: YYY

## Appendix C

*GPT-4o Example Output*

*Let's begin by analyzing the two psychometric items in detail:*

*\*\*Item 1: I do too little work\*\**

*\*\*Item 2: I feel that others misunderstand me\*\**

*### Step-by-Step Analysis*

*#### Step 1: Identifying Key Psychological Constructs*

***Item 1: I do too little work***

*- This item likely taps into constructs such as:*

  *- **Work Ethic**: The belief in the importance of hard work and diligence.*

  *- **Self-Efficacy**: One's belief in their ability to succeed in specific situations.*

  *- **Procrastination**: The action of delaying or postponing tasks.*

  *- **Motivation**: The reason or reasons one has for acting or behaving in a particular way.*

***Item 2: I feel that others misunderstand me***

*- This item likely taps into constructs such as:*

  *- **Interpersonal Perception**: How one perceives others' understanding of them.*

  *- **Social Anxiety**: The fear of social situations that involve interaction with other people.*

  *- **Self-Esteem**: One's overall subjective emotional evaluation of their own worth.*

  *- **Communication Skills**: The ability to convey information to another effectively and efficiently.*

*#### Step 2: Drawing from Empirical Research*

*- **Work Ethic and Interpersonal Perception**: Research suggests that individuals with a strong work ethic may feel misunderstood if they perceive others as not recognizing their efforts. However, this is not a direct relationship and may be mediated by other factors such as communication skills and self-esteem.*

*- **Self-Efficacy and Social Anxiety**: Low self-efficacy can be associated with higher social anxiety, which in turn can lead to feelings of being misunderstood.*

*- **Procrastination and Self-Esteem**: Procrastination can lead to lower self-esteem, which might make individuals feel misunderstood by others.*

*- **Motivation and Communication Skills**: Low motivation can affect one's ability to communicate effectively, potentially leading to feelings of being misunderstood.*

*#### Step 3: Analyzing Past Research Correlations*

*- **Work Ethic and Social Anxiety**: Studies have shown a weak negative correlation between work ethic and social anxiety (e.g., -0.10 to -0.20).*

*- **Self-Efficacy and Self-Esteem**: There is a moderate positive correlation between self-efficacy and self-esteem (e.g., 0.30 to 0.50).*

*- **Procrastination and Self-Esteem**: There is a moderate negative correlation between procrastination and self-esteem (e.g., -0.30 to -0.40).*

*- **Motivation and Communication Skills**: There is a weak to moderate positive correlation between motivation and communication skills (e.g., 0.20 to 0.30).*

*#### Step 4: Considering Whether There is a Correlation*

*Given the constructs involved, it is plausible that there is a correlation between the two items, but it is likely to be weak to moderate. The relationship is not direct but mediated by factors such as self-esteem, social anxiety, and communication skills.*

*### Further Assessment*

#### Fundamental Psychological Concepts

- **Self-Esteem**: Both items can be indirectly related through self-esteem. Low self-esteem can lead to both feeling misunderstood and perceiving oneself as doing too little work.

- **Social Anxiety**: Social anxiety can lead to feelings of being misunderstood and can also affect one's work output.

#### More Complex Theories

- **Cognitive Dissonance**: Individuals who feel they do too little work might experience cognitive dissonance, leading them to believe that others misunderstand them to justify their behavior.

- **Attribution Theory**: People might attribute their lack of work to external factors, leading to feelings of being misunderstood.

### Mapping Out Reasoning Paths

1. **Path 1**: Low self-esteem ➔ Feelings of doing too little work ➔ Feelings of being misunderstood.

2. **Path 2**: High social anxiety ➔ Feelings of being misunderstood ➔ Perception of doing too little work.

3. **Path 3**: Poor communication skills ➔ Feelings of being misunderstood ➔ Low motivation ➔ Doing too little work.

### Self-Consistency Evaluation

Across different lines of reasoning, the most consistent outcome is a weak to moderate correlation. The constructs involved (self-esteem, social anxiety, communication skills) suggest a relationship, but not a strong one.

### Considering Reasons Against the Analysis

1. **Individual Differences**: Variability in individual experiences and personality traits might weaken the correlation.

2. **Measurement Error**: The way the items are phrased might lead to different interpretations, affecting the correlation.

3. **Contextual Factors**: External factors such as work environment and social context might influence the responses, reducing the correlation.

### Final Estimated Correlation

After thorough analysis, the most plausible estimate of the correlation between the two items is:

**Predicted Correlation: 0.15**

This estimate reflects a weak positive correlation, consistent with the constructs and empirical research discussed.

*Claude 3 Opus Example Output*

Okay, let's think through this step-by-step to estimate the correlation between "I do too little work" and "I feel that others misunderstand me":

a) Key psychological constructs:

- Item 1 relates to work ethic, motivation, and productivity

- Item 2 relates to feelings of being misunderstood, social disconnection, and interpersonal relationships

b) Past research on the relationship between these constructs:

*- Studies have found that social support and positive relationships at work are associated with higher job satisfaction and productivity (e.g. Viswesvaran et al., 1999)*

*- However, the specific link between feeling misunderstood by others and doing too little work has not been well-established empirically to my knowledge*

*c) Statistical correlations from past research:*

*- Meta-analyses show the correlation between social support and job performance is typically small to moderate, around r = 0.25 (Viswesvaran et al., 1999)*

*- But again, the specific variables here have not been directly correlated in published research I'm aware of*

*d) Considering if there is any correlation at all:*

*- There could plausibly be a small correlation, if feeling misunderstood leads to lower motivation and productivity at work*

*- However, many other factors influence work ethic and productivity beyond just interpersonal relationships*

*- It's also very possible there is no meaningful correlation between these two specific variables*

*Progressing from fundamental to more complex concepts:*

*- At a basic level, interpersonal relationships and work performance are separate domains*

*- More complex theories propose that social factors can influence work outcomes to some degree*

*- However, the specific subjective feeling of being misunderstood may not necessarily translate directly to objective measures of doing too little work*

*- Many individual and contextual factors likely moderate any relationship between these constructs*

*Mapping out different lines of reasoning:*

*- One perspective is that feeling misunderstood could reduce motivation and lead to underperformance at work, resulting in a positive correlation*

*- Alternatively, people could compartmentalize social and work life, resulting in no correlation*

*- A third view is that those who feel misunderstood may overcompensate by working harder to prove themselves, resulting in a negative correlation*

*- The most likely outcome based on past research is probably a small positive correlation or no meaningful correlation*

*Evaluating consistency and cross-validating:*

*- The "no meaningful correlation" outcome seems most consistent across different approaches*

*- Empirical research has not established a clear link between these specific variables*

*- Conceptually, while social factors can influence work, this specific link is tenuous*

*- Most effects in this area of research tend to be small to negligible*

*Considering reasons I might be mistaken:*

*- There could be certain populations or contexts where this correlation is stronger than I'm assuming*

*- I may be overestimating the tendency for people to compartmentalize interpersonal and work domains*

*- Unmeasured third variables could be causing a spurious correlation between these two items*

*Final estimate:*

*Based on a thorough analysis considering past research, psychological theories, and logical reasoning, my final estimate is that the correlation between "I do too little work" and "I feel that others misunderstand me" is likely to be small to negligible in a broad US population. Interpersonal and productivity domains are often separate, meta-analyses show social variables have only small effects on work performance, and no research has established a clear empirical link between these specific constructs. Accounting for self-consistency and potential errors in my reasoning, I predict the most plausible correlation is:*

*Predicted Correlation: 0.05*